\documentclass[aps,prc,showpacs,groupedaddress]{revtex4-1}

\usepackage{graphicx} 
\usepackage{dcolumn}  
\usepackage{bm}       
\usepackage{amsmath}
\begin{document}
\newcommand{\vv}[1]{{$\bf {#1}$}}
\newcommand{\ul}[1]{\underline{#1}}
\newcommand{\vvm}[1]{{\bf {#1}}}

\title{Elastic Scattering of $^6$He based on a Cluster Description}

\author{S.P. Weppner}
\email[1]{weppnesp@eckerd.edu}

\altaffiliation{Permanent Address: 
Natural Sciences, Eckerd College, St. Petersburg, FL 33711}

\author{Ch. Elster}
\email[2]{elster@ohio.edu}
\affiliation{Institute of Nuclear and Particle Physics,  and
Department of Physics and Astronomy,  Ohio University, Athens, OH 45701
}

\date{\today}

\begin{abstract}
Elastic scattering observables (differential cross section and analyzing power) are
calculated for the reaction $^6$He(p,p)$^6$He at projectile energies starting at
71~MeV/nucleon. The optical potential needed to describe the reaction is derived
by describing $^6$He in terms of a $^4$He-core and two neutrons. The Watson first order
multiple scattering ansatz is extended to accommodate the internal dynamics of a 
composite cluster model for the $^6$He nucleus scattering from a nucleon
projectile. The calculations are compared with the recent experiments at 
the projectile energy of 71~MeV/nucleon. In addition, 
differential cross sections and analyzing powers are calculated at selected
higher energies.
\end{abstract}


\maketitle

\section{Introduction}

Traditionally, differential cross sections and spin observables played an important
role in either determining the parameters in phenomenological optical models for
proton-nucleus (p-A) scattering or in testing the accuracy and validity
of microscopic models thereof. Specifically, elastic scattering of protons and
neutrons from stable nuclei lead to a large body of work on optical
potentials in which the two-nucleon interaction and the density matrix of the
nucleus were taken as input to ab-initio calculations of first order optical
potentials, either in a KMT or Watson expansion of the multiple scattering series,
for which the primary goal was a deeper understanding of the reaction mechanism.

For exotic nuclei the theoretical emphasis is somewhat shifted. A goal 
of  microscopic reaction theory is a construction of the scattering observables
based on well-defined dynamical and structure quantities in order to, for example,
examine structural sensitivities. 
Investigating the
structure of halo nuclei, specifically $^6$He, has already inspired a large body of
work including few-body models~\cite{Zhukov:1993aw}, Green's Function Monte Carlo (GFMC)
methods~\cite{Pudliner:1997ck}, and 
no-core shell model calculations~\cite{Navratil:2004dp}, so that ground state
properties of $^6$He appear to be quite well understood.  

Recently, elastic scattering of $^6$He off a polarized proton target has been measured 
for the first time at an energy of 71~MeV/nucleon~\cite{Uesaka:2010mm,Hatano:2005}.
The experiment finds that the analyzing power becomes negative around 50$^o$
which is not predicted by simple folding models for the optical
potentials~\cite{Weppner:2000fi,Gupta:2000bu}. The same calculations  
nevertheless describe the
differential cross section at this energy reasonably well.
This apparent ``$A_y$ problem" conveys the inadequacy of 
using the same
models which describe p-A scattering from stable nuclei 
for reactions involving halo nuclei. The obvious difference is the nuclear
structure. While the typical stable nuclei for which folding models are very
successful are mostly spherical, $^6$He can be understood in few-body models 
as a three-body system consisting of 2 neutrons ($n$) and a $^4$He core. 
Implementing this three-body structure in a cluster model, specifically in a 
reaction calculation for proton ($p$) scattering off $^6$He, 
was pioneered in Ref.~\cite{Crespo:2006vg} 
for calculating the reaction $p+^6$He at 717~MeV/nucleon, 
an energy at which the authors could
employ the Born approximation. Based on the KMT formulation for the optical potential
and more realistic wave functions for $^6$He, differential cross sections and
analyzing powers were calculated in Ref.~\cite{Crespo:2007zz} at 297~MeV/per
nucleon. For describing the differential cross section and the analyzing power at
71~MeV/nucleon, Ref.~\cite{Uesaka:2010mm} suggested a ``cluster-folding" calculation
having an explicit $\alpha$-core described by a phenomenological $p+^4$He optical
potential. 

In this work we want to extend the development of Ref.~\cite{Crespo:2006vg} by
incorporating the cluster structure in an optical potential for the reaction
$p+^6$He in such a way that the transition amplitude can be iterated to all orders
(non-Born approximation).
Our derivation of the optical potential is based on the Watson formulation for the
multiple scattering theory, which not only allows to treat proton and neutron
contributions to the structure separately~\cite{Chinn:1993zza}, but also lends
itself naturally to taking into account the contributions of the $\alpha$-core and
the two neutrons. The construction of an optical potential in which
the separate contributions from the clusters are treated in a consistent fashion 
is achieved.

This article is organized as follows. In Section II we first present a short
summary of the Watson optical potential for stable nuclei, and then extend this
paradigm to the $^6$He nucleus consisting of an $\alpha$-core and two neutrons. 
In Section III we present our calculation for $^6$He$+p$ at 71~MeV/nucleon
as well as at several higher energies and discuss their implications. Our
conclusions are presented in Section IV. Three Appendices are devoted to 
the explicit derivation of the first order optical potential, the
transformations between the different coordinate systems used in our 
calculations,
and the calculation of the correlation densities between the clusters.


\section{The Folding Cluster Model}

In order to derive a cluster ansatz for the target (projectile), and show how it
can consistently be
incorporated into a folding approach for the optical potential, we will for the
convenience of the reader give a summary of the essential ingredients and underlying
assumptions.

\subsection{The Watson Optical Potential for Single Scattering}

Let $H=H_0+V$ be the Hamiltonian for the nucleon-nucleus system (A+1 body system),
where the interaction $V=\sum_i^A v_{0i}$ consists of all two-nucleon interactions 
$v_{0i}$ between the projectile (``$0$'') and a target nucleon (``$i$''). The free
Hamiltonian is given by $H_0 = h_0 + H_A$, where $h_0$ describes the kinetic energy
of the projectile, while the target Hamiltonian $H_A$ satisfies $H_A |\Psi_A\rangle =
E_A|\Psi_A\rangle$, with $|\Psi_A\rangle$ being the ground state of the target.

The transition amplitude for the scattering of the projectile from the target is then
given by a Lippmann-Schwinger equation, $T= V+ VG_0 T$, where the propagator $G_0$ is an
(A+1) body operator given by
\begin{equation}
G_0(E) = (E - h_0 - H_A + i\varepsilon)^{-1},
\label{eq:2.1}
\end{equation}
with $E$ being the total energy of the system. One way to tackle the many-body
scattering problem is the spectator expansion~\cite{Siciliano:1977zz}, which writes
the transition amplitude as
$T=\sum_{i=1}^A T_{0i}$, so that 
\begin{equation}
T_{0i}=v_{0i}+v_{0i}G_0(E)T.
\label{eq:2.2}
\end{equation}
This allows the sum of all interactions between projectile `$0$' and nucleon `$i$' by
a formal reordering of the multiple scattering series according to Watson,
\begin{eqnarray}
T_{0i}= {\hat \tau}_{0i} +{\hat \tau}_{0i}G_0(E)\sum_{j\ne i}T_{0j}, 
\label{eq:2.3}
\end{eqnarray}
where
\begin{equation} 
{\hat \tau}_{0i}=v_{0i}+v_{0i}G_0(E){\hat \tau}_{0i}.\label{eq:2.4}
\end{equation} 
The term ${\hat \tau}_{0i}$ of Eq.~(\ref{eq:2.4}) only
involves  the interaction between pairs, namely particles `$0$' and `$i$', whereas 
the propagator $G_0(E)$ is still an (A+1)-body operator. 
The multiple scattering series of Eq.~(\ref{eq:2.3}) can directly serve as starting 
point for constructing the transition amplitude for elastic scattering as shown in
Refs.~\cite{Crespo:2006vg,Crespo:2007zz,Crespo:2001nn}.

When focusing on elastic scattering, the projection $P$ onto the
ground state $|\Phi_A\rangle$ is introduced so that $[G_0(E),P]=0$. Here we define
$P=\frac{|\Phi_A \rangle \langle \Phi_A|}{\langle \Phi_A| \Phi_A \rangle }$ 
and $P+Q={\bf 1}$, where $Q$ projects onto the orthogonal space.
This allows the separation of the transition amplitude into two pieces,
\begin{eqnarray}
T&=&U+UG_0(E)PT \cr
U&=&V+VG_0(E)QU, 
\label{eq:2.5}
\end{eqnarray}
with $U$ being the optical potential operator.  The transition operator 
for elastic scattering
may then be defined as $T_{el}=PTP$, so that 
\begin{equation}
T_{el}=PUP + PUPG_0(E)PT_{el}
\label{eq:2.6}
\end{equation}
is a one-body integral equation. 
Of course, it requires the knowledge of $PUP$, which has to
contain the complete information about the many-body character of the
problem.
The formulations for the transition matrix for elastic scattering, given in 
Eqs.~(\ref{eq:2.3}) and (\ref{eq:2.5}), are equivalent though truncations
in the expansions are not.
 
The first order term of $U$ can be defined as
\begin{equation}
U=\sum_{i=1}^A U_i\approx \sum_{i=1}^A \tau_{0i}, \label{eq:2.7}
\end{equation}
with
\begin{equation}
{\tau_{0i}} = v_{0i} + v_{0i} G_0(E) Q {\tau_{0i}}. \label{eq:2.8}
\end{equation}
Because of the appearance of the projection operator $Q$ in 
Eq.~(\ref{eq:2.7}), the quantity $\tau_{0i}$ can not yet be 
related to a two nucleon interaction.
Defining a transition operator ${\hat \tau}_{0i}$, according to Eq.~(\ref{eq:2.4}),
allows the explicit relation to $\tau_{0i}$:
\begin{eqnarray}
{\hat\tau}_{0i}&=& v_{0i} + v_{0i} G_0(E)\left[P+Q \right] {\hat\tau}_{0i} \nonumber \\
&=& \tau_{0i}+\tau_{0i}G_0(E)P{\hat\tau}_{0i},
\label{eq:2.9}
\end{eqnarray}
so that one obtains the exact relations~\cite{Chinn:1993zza}
\begin{equation}
\tau_{0i} ={\hat\tau}_{0i} - \tau_{0i} G_0(E) P {\hat\tau}_{0i} =
{\hat\tau}_{0i} - {\hat\tau_{0i}} G_0(E) P \tau_{0i}.
\label{eq:2.10}
\end{equation}
Taking into account the iso-spin character of the target nucleons instead 
of just summing over $A$ nucleons can be easily done by splitting 
 Eq.~(\ref{eq:2.7}) into 
two parts under the assumption that the projectile `0' is
a proton, Eq.~({\ref{eq:2.7}}) becomes:
\begin{equation}
U_p=\sum_{i=1}^Z \tau_{0i}^{pp} + \sum_{i=1}^N \tau_{0i}^{np}
\equiv U_p^Z+U_p^N, \label{eq:2.11}
\end{equation}
where the integral Eq.~({\ref{eq:2.10}}) has to be solved 
separately for proton-proton (pp) and neutron-proton (np) interactions. 
This clearly indicates 
that the optical potential for the scattering of a proton ($U_p$)
from a target nucleus differs from the optical potential
for the scattering of a neutron ($U_n$) from the same target. Moreover, and more
important for the present considerations, the folding with the proton and 
neutron density matrices is cleanly separated, which is not the case if
one uses an optical potential in the formulation of 
Kerman-McManus-Thaler (KMT)~\cite{KMT}. A numerical study for $p+^{11}Li$
between a KMT formulation and Watson expansion of Eq.~(\ref{eq:2.3}) has been
carried out in Ref.~\cite{Crespo:2001nn} with the finding that truncations 
at the same order of the series though being similar at small momentum transfer
show differences at the higher momentum transfers. This should not be surprising 
when considering the nonlinear relation of the free two-nucleon t-matrix, 
Eq.~(\ref{eq:2.10}), to the quantity $\tau_{0i}$ entering the optical potential
which is the driving term of the final scattering integral equation, 
Eq.~(\ref{eq:2.6}). For the $n+d$ system, for which exact solutions of the
Faddeev equations exist, optical potentials for elastic scattering were
constructed in~\cite{Kuros:1997zz} indicating that first order approximations
are only valid for smaller momentum transfers.

The propagator of Eq.~(\ref{eq:2.1})
needs to be examined in more detail since it still is an
(A+1)-body operator. Only if the target Hamiltonian $H_A$
is approximated by a c-number, $G_0(E)$ becomes a one-body
operator and Eq.~(\ref{eq:2.4}) can be identified with the 
free nucleon-nucleon (NN) t-matrix at an appropriate energy E.
Reducing $H_A$ to a c-number is the standard impulse
approximation (or extreme closure approximation)
which has been widely used throughout the literature. 
The impulse approximation is believed to be a reasonable
approximation in intermediate energy nuclear physics, i.e.
in an energy regime where the kinetic energy of the projectile 
is large compared to excitation energies of the target, however
the validity of this assumption has to be tested for individual 
cases under consideration.

\subsection{First Order Folding Optical Potential}

\begin{figure}[htbp]
\begin{center}
\includegraphics[scale=.45]{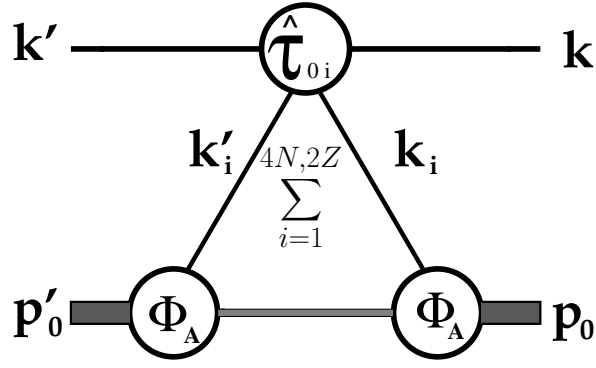}
\caption{Diagram for the standard optical potential matrix 
element for the single-scattering approximation.
\label{fig1}
}
\end{center}
\end{figure}

In this section we will give the explicit expressions for
the `traditional'  first order Watson optical potential. Starting from those expressions
will lead in a straightforward fashion to an optical potential where
the nucleus is treated as a composite of clusters. Based on Eqs.~(\ref{eq:2.6}) and 
(\ref{eq:2.7}) the first order optical potential as function of external momenta
\vv k and $\vvm k'$ is given by
\begin{equation}
\langle \vvm k' | \langle \phi_A|PUP|\phi_A\rangle
\vvm k\rangle\equiv U_{el}(\vvm k',\vvm k)=\sum_{i=N,P}\left
\langle \vvm k' | \langle \phi_A|\hat\tau_{0i}({\cal{E}})|\phi_A\rangle
\vvm k\right\rangle \equiv \langle {\hat \tau}_{0i}\rangle,\label{eq:2.8.1}
\end{equation}
where ${\cal{E}}$ is the energy of the system.
The summation over $i$ indicates that one has to sum over $N$ neutrons and $Z$
protons.
The structure of Eq.~({\ref{eq:2.8.1}}) is graphically
indicated by Fig.~\ref{fig1}, $\vvm k_i$ and $\vvm k_i'$ 
are internal variables of the struck target nucleon, 
$\vvm p_0$ and $\vvm p_0'$ are external target variables.
In the following derivation, we will sum over $A$ nucleons for clarity of presentation.
However it should be emphasized that in practice a sum over $N$ neutrons and $Z$
protons, as indicated in Eq.~(\ref{eq:2.10}), is done. 
The energy of the $\hat\tau_{0i}$
operator, $\cal{E}$, is a dynamical variable which depends on the total 
energy of the system and the energy of the spectator 
\cite{Elster:1997as} similarly to a three-body problem.
In this work the common
approximation to fix it at half the laboratory energy will be used.

We consider a frame of reference in which
${\bf k_0}$ is the momentum of the projectile, 
and ${\bf p_0}$ is the momentum of the target. 
The individual nucleons inside the target have 
momenta $(\vvm k_1,\vvm k_2,...\vvm k_A)$. Thus
\begin{equation}
\langle \vvm k_1\vvm k_2\vvm k_3\vvm k_4...\vvm k_A|\phi_A\rangle=
\delta(\vvm k_1+\vvm k_2+\vvm k_3 +\vvm k_4...+\vvm k_A-\vvm p_0)
\langle {\bf\zeta_1\zeta_2\zeta_3\zeta_4...\zeta_{A-1}}
|\phi_A\rangle,
\label{eq:2.8.1a}
\end{equation}
where
the delta function determines the conservation of the absolute momentum for the 
center of mass  (c.m.) frame of the nucleus, and $\vvm \zeta_i$ represent 
the relative momenta of the individual nucleons in the target.
The reason for using relative momenta is that
the wave functions are naturally  expressed in such a basis to manifestly
express Galilean invariance,
so any input into the theory will be in terms of the internal momenta
${\bf \zeta_i}$ and not $\vvm k$. 
In first order, which is the concern of this work, we only need the momentum
of the struck nucleon,
namely $\vvm k_{1}={\bf \zeta_1} + \frac{\vvm p}{A}$.
Changing  integration variables from absolute to 
relative momenta, only the 
single particle density matrices, $\rho({\bf\zeta_1',\zeta_1})$, are employed.
They are used to describe the dependence of one particle's  
relative motion to the remaining (A-1)-core,
\begin{equation} 
\rho({\bf \zeta_1',\zeta_1}) 
 \equiv \int\prod_{l=2}^{A-1}d{\bf \zeta_l'}
 \int\prod_{j=2}^{A-1}d{\bf \zeta_j} 
\langle \phi_A |\vvm\zeta_1'\vvm\zeta_2'\vvm\zeta_3'\vvm\zeta_4'...
\vvm\zeta_{A-1}'\rangle\;
\langle \vvm\zeta_1\vvm\zeta_2\vvm\zeta_3\vvm\zeta_4...\vvm\zeta_{A-1}
|\phi_A\rangle.\label{eq:2.8.1b}
\end{equation}
The NN $\hat\tau_{01}$-matrix can 
always be written in relative coordinates as
\begin{equation}
\langle\vvm q_0'\vvm q_1'|\hat\tau_{01}({\cal E})
|\vvm q_0 \vvm q_1\rangle =
\delta(\vvm q_0'+\vvm q_1'-\vvm q_0-\vvm q_1) \left\langle
\frac{1}{2}(\vvm q_0'-\vvm q_1') \big|
\hat\tau_{01}({\cal E})\big| \frac{1}{2}(\vvm q_0-\vvm q_1)\right\rangle,
\label{eq:2.8.1c}
\end{equation}
where the $\delta$-function indicates the momentum conservation of the 
two-nucleon pair.
This leads to the following expression for the optical potential of 
Eq.~(\ref{eq:2.8.1}),
\begin{eqnarray}
\langle\hat\tau_{01}\rangle &=&
\;\int d{\bf\zeta_1'}
\;\int d {\bf\zeta_1}\;\delta(\vvm k' + \vvm p_0' - \vvm k -
\vvm p_0)\; \nonumber \\
&&\left\langle\frac{1}{2}\left(\vvm k'-{\bf\zeta_1'}
-\frac{\vvm p_0'}{A}\right) \Big|{\hat\tau}_{01}({\cal{E}}) \Big|
\frac{1}{2}\left(\vvm k- {\bf\zeta_1}-\frac{\vvm p_0}{A}\right)\right\rangle\;
\rho({\bf\zeta_1'},{\bf\zeta_1}) \nonumber \\
&&\delta\left(\zeta_1'-\zeta_1- \frac{A-1}{A}(\vvm k-\vvm k')\right),
\label{eq:2.8.8b}
\end{eqnarray}
which explicitly gives the following relation
\begin{equation}
{\bf\zeta_1'}- {\bf\zeta_1}=
\frac{A-1}{A}(\vvm k- \vvm k'),
\end{equation}
relating the relative variables ${\bf \zeta}$ directly
to the external variables. For convenience, in the 
practical calculation, the definition
\begin{eqnarray}
\vvm K &\equiv& \frac{1}{2}(\vvm k'+\vvm k) \nonumber \\
\vvm q &\equiv& \vvm k'-\vvm k =
\frac{A}{A-1}({\bf\zeta_1}- {\bf\zeta_1'}) \nonumber \\
\vvm P& \equiv& \frac{1}{2}({\bf\zeta_1'}+{\bf\zeta_1}),\label{eq:2.8.8a}
\end{eqnarray}
where \vv q is the momentum transfer and \vv K is orthogonal 
to it, is used.
After a series of variable transformations, which are outlined
in Appendix~\ref{appendix0},  one obtains for folding 
single-particle optical potential
\begin{eqnarray}
\lefteqn{U_{el}(\vvm q, \vvm K) = } \cr
&&\sum_{i=n,p}
\int d \vvm P \hat\tau_{0i}\left(\vvm q ,
\frac{1}{2}\left(\frac{A+1}{A}\vvm K - \vvm P\right),{\cal{E}}\right)
\;\rho_i\left(\vvm P-\frac{A-1}{2A}\vvm q
,\vvm P+\frac{A-1}{2A}\vvm q \right). 
\label{eq:2.8.11}
\end{eqnarray}
The above expression shows that one has to carry out a three-dimensional integration over
the NN t-matrix and the nuclear density matrix. We are employing Monte Carlo methods for
the actual computation. 


\subsection{Cluster Model for the Target Nucleus}

Halo nuclei exhibit the distinctive feature of a usually tightly bound core
and loosely bound valence nucleons. For exploring bound state properties a
cluster model has been quite successful (see e.g.
~\cite{Zhukov:1993aw,Ghovanlou:1974zza}).
Here we propose to employ a cluster model for the optical potential of $^6$He,
and thus view $^6$He as a cluster of a tightly bound $\alpha$-particle and two valence
neutrons. As explained in the previous sections, the first order folding optical
potential uses the single nucleon density as a basic ingredient. That means, in the
standard formulation of the optical potential, one deals with one active nucleon at a
time, and then sums over all active nuclei.
This paradigm can be
naturally extended to a cluster model. The only additional consideration which
needs to be taken into account is that now there is an intrinsic motion between the
different pieces of the cluster. In order to accommodate this
let us define a Jacobi momentum ${\bf p_j}$, 
representing the relative momentum of the active
cluster and the spectator clusters, as
\begin{equation}
{\vvm p_j}_i = \frac{1}{A}({A_s}_i \vvm p_i - A_i \vvm {p_s}_i),\label{eq:c1}
\end{equation}
where the index $i$ characterizes a
particular cluster and the index $s_i$ represents the spectators 
for the $i$th cluster. Since 
this is a Jacobi momentum it is invariant in all frames. The 
underlying belief is that there is a strongly peaked 
active cluster momentum,
centered about ${\bf p_i}$, which is different from the other spectator 
cluster momentum, ${\bf p_s}_i$. If the nucleons are assumed to be largely
independent of each other, as in a single particle picture,  then
${\vvm p_j}_i$ = 0, and the single-particle optical potential re-emerges. 

With these additional momenta we can define a correlation
density similar to the traditional density of Eq.~(\ref{eq:2.8.1b}):
\begin{equation} 
\rho_{corr}({\vvm p_j}_1,{\vvm p_j}^\prime_1) 
 \equiv \int\prod_{l=2}^{N_c}d{{\vvm p_j}^\prime_l}
 \int\prod_{m=2}^{N_c}d{{\vvm p_j}_m} 
\langle \phi_A |\vvm {p_j}^\prime_1\vvm {p_j}^\prime_2 ...
\vvm {p_j}^\prime_{N_c}\rangle\;
\langle \vvm {p_j}_1\vvm {p_j}_2 ...
\vvm {p_j}_{N_c} |\phi_A\rangle,
\label{eq:c2}
\end{equation}
where $N_c$ is the number of total clusters in the target. 
For the $^6$He system studied here
$N_c=3$.
The particular view in this model is  that  
nucleons within a specific cluster move with the same c.m. momentum,
which in turn is correlated with the momentum of the
spectator clusters. This product of 
densities must conserve the overall momentum in the intrinsic frame of the nucleus.
It further follows that the sum of all relative momenta in the cluster must be zero.
In this work, we choose to work with the three-body cluster orbital shell model
approximation (COSMA) density~\cite{Zhukov:1994zz,Zhukov:1991}.

\begin{figure}[htbp]
\begin{center}
\includegraphics[scale=.35]{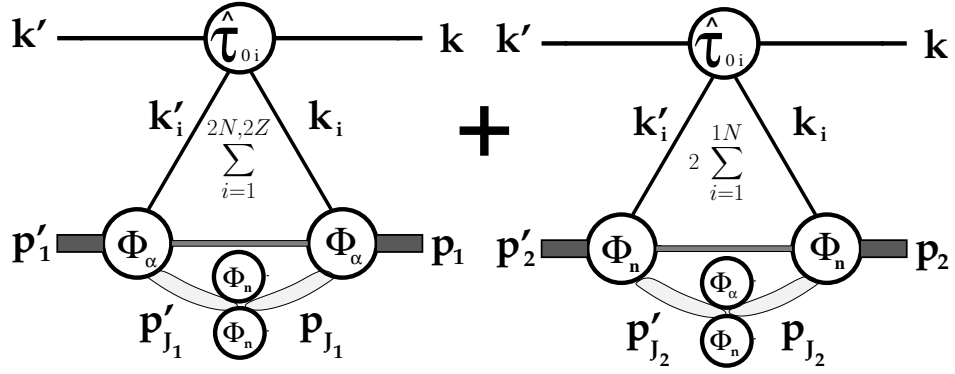}
\caption{Diagram for the cluster optical potential for $^6$He based on the
single-scattering approximation.
\label{fig2}
}
\end{center}
\end{figure}

In order to match the choice of momenta to those of the optical potential of the previous
subsection, we define
\begin{equation}
{\vvm {\cal P}_j}_i = \frac{{\vvm p_j}_i+ {\vvm p_j}^\prime_i}{2},\label{eq:c3}
\end{equation}
which is similar to the definitions of $\vvm P$ and $\vvm K$ defined 
in Eq.~(\ref{eq:2.8.8a}).
Thus, the new cluster optical model 
provides an additional sum over the number of clusters
\begin{eqnarray}
U_{el}(\vvm q, \vvm K) &=&\sum_{c=1,N_c}\;\sum_{i=n_c,p_c}
\int d \vvm P \; d{\vvm {\cal P}_j}_c \; \rho_{corr}
({\vvm {\cal P}_j}_c) \nonumber \\
&&\hat\tau_{0i}\left(\vvm q , \frac{1}{2}\left(\frac{A+1}{A}\vvm K - \vvm P\right),
{\cal{E}}\right)
\;{{\rho}_c}_i \left(\vvm P-\frac{A-1}{2A}\vvm q
,\vvm P+\frac{A-1}{2A}\vvm q \right), 
\label{eq:2.8.11b}
\end{eqnarray}
where each cluster now  defines its own optical potential. 
With this, the optical potential for  
$^6$He consists of two pieces as indicated in Fig.~\ref{fig2}, 
an optical potential for the $\alpha$-core and one for
each of the two neutrons, both linked by the correlation density between the clusters,
\begin{eqnarray}
\lefteqn{^{^6\rm{He}} U_{el} (\vvm q, \vvm K) = U_\alpha + 2 U_{n} =} \cr
&&\sum_{i=N,P} \int d \vvm P \; d{\vvm {\cal P}_j}_\alpha 
\; \rho_{corr}({\vvm {\cal P}_j}_\alpha)
\;\hat\tau_{0i}\left(\vvm q ,
\frac{1}{2}\left(\frac{A+1}{A}\vvm K - \vvm P\right),{\cal{E}}\right)
\;{{\rho}_\alpha}_i\left(\vvm P-\frac{A-1}{2A}\vvm q
,\vvm P+\frac{A-1}{2A}\vvm q \right) \nonumber \\
&&+ 2 
\int d \vvm P \; d{\vvm {\cal P}_j}_n
\;\rho_{corr}({\vvm {\cal P}_j}_n)
\;\hat\tau_{0i}\left(\vvm q , \frac{1}{2}(\frac{A+1}{A}\vvm K - \vvm
P),{\cal{E}}\right)
\;{\rho}_n\left(\vvm P-\frac{A-1}{2A}\vvm q,\vvm P+\frac{A-1}{2A}\vvm q \right).
\label{eq:2.8.11c}
\end{eqnarray}
The cluster optical potential involves now a six-dimensional integration, which we again
carry out with Monte Carlo methods.
In addition, care must be taken to evaluate each cluster in the c.m. frame
between projectile and cluster, since the scattering takes place
in intrinsically different frames as dictated by the variable 
$\vvm {\cal P}_j$. However, the final optical potential for the nucleus
$^6$He  must be evaluated
in the c.m. frame of the target+projectile system. 
Thus, the initial step in the  calculation involves first the evaluation of each
cluster optical potential separately, and 
then boosting each to the target+projectile system. The explicit details of these
transformations are given in Appendix~\ref{appendixa}.

Furthermore, it is important to note that each cluster optical potential 
contains the correlation density.
Through its variable $\vvm {\cal P}_j$ the momenta of the active particles are 
constrained. An explicit derivation of the correlation density for the $^6$He nucleus
is given in Appendix~\ref{appendixb}.


\section{Results and Discussion}

In this section we evaluate the differential cross section and the analyzing power for
elastic scattering of $^4$He and $^6$He using non-local optical potentials in first order in
the Watson multiple scattering expansion. Specifically, we want to test the influence of the
cluster formulation presented in Section II.C on 
those observables compared to single-particle
optical potentials of the same order in the multiple scattering expansion. 
We start with considering  the recent experimental
data for $^6$He at 71~MeV/nucleon~\cite{Uesaka:2010mm}, then we continue our investigation
at slightly higher energies in order to gain some insight on the behavior of the elastic
scattering observables as function of projectile energy.

As a  
nucleon-nucleon (NN)
interaction we use the nonlocal Nijm-I potential of the Nijmegen group~\cite{Stoks:1993tb},
which describes the NN data below 350~MeV laboratory energy with a $\chi^2 \approx 1$. 
For the density matrix of $^6$He the 
COSMA density of Refs.~\cite{Zhukov:1994zz,Zhukov:1991} is used. 
This density consists of harmonic
oscillator wave functions for the s- and p-shell. The parameters are fitted such that 
$^6$He has a charge radius of 1.77~fm and a matter radius of 2.57~fm. 

In addition the elastic scattering observables for proton scattering off
$^4$He is calculated. The proton and neutron density matrix for $^4$He is obtained
from a microscopic Hartree-Fock-Bogoliubov (HFB) calculation of Ref.~\cite{HFB}, which uses
the Gogny D1S finite range effective interaction~\cite{Gogny}.
The HFB calculation produces a mean-field potential, which in turn is used
to modify the free NN interaction inside the nucleus~\cite{Chinn:1995qn,Chinn:1993zz}.
This modification of the free NN t-matrix has proved to be important in the description of
closed shell nuclei at projectile energies below 150~MeV. 

\begin{figure}[htbp]
\begin{center}
\includegraphics[scale=.75]{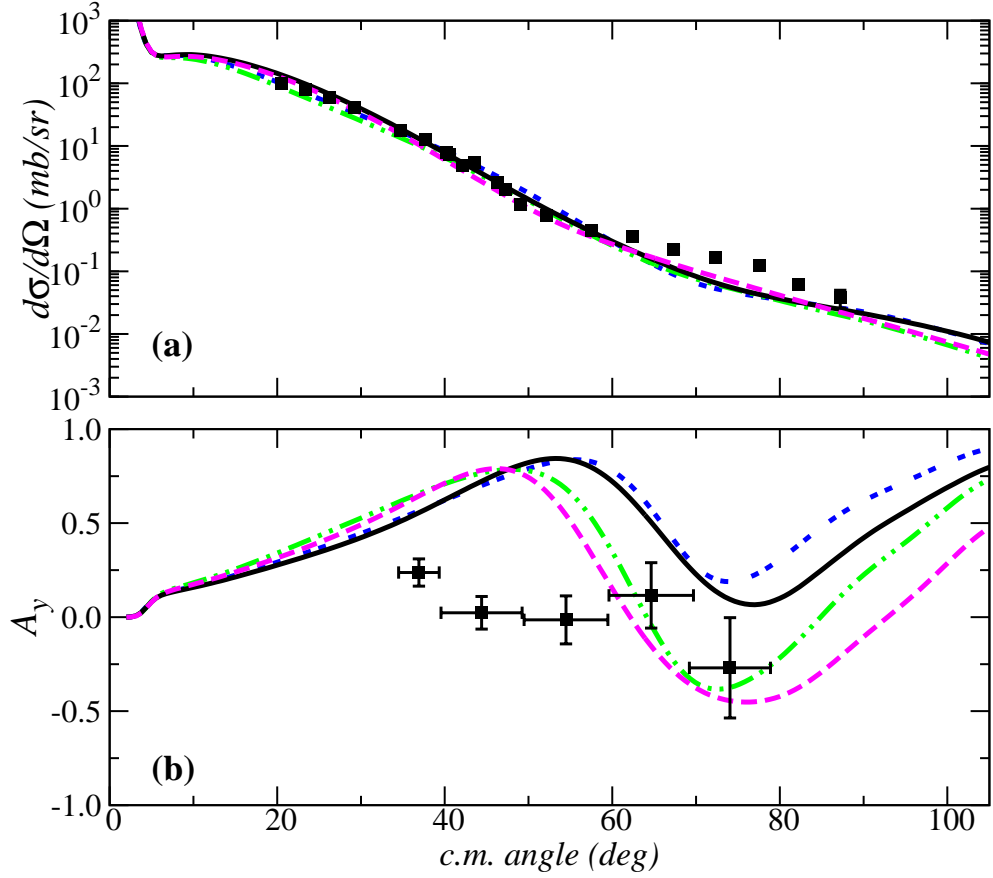}
\caption{(color online) The angular distribution of the differential cross
  section,$\frac{d\sigma}{d\Omega}$, (upper panel (a)) and the analyzing power, $A_y$, 
(lower panel (b)) for
  elastic scattering $^{6}$He at projectile energy 71~MeV/nucleon as function of
the c.m. angle.
The calculations are performed with optical potential
  obtained from the Nijmegen I potential~\cite{Stoks:1993tb} for the NN interaction.
All optical potential are folding, non-local optical potentials described in the
text.
The solid line (black) represents the calculation based on a single-particle optical
potential employing the COSMA density of Ref.~\cite{Zhukov:1994zz}. For the short-dashed line
(blue) the cluster ansatz together with the COSMA density is used. The dash-double-dot line
(green) represents a calculation based on the cluster formulation, however 
the NN t-matrix for the core optical potential is modified by a mean field obtained from
a HFB~\cite{HFB,Gogny} calculation.
The short dashed line (pink)
represents the same calculation, but neglects correlation of the clusters. 
The data are taken from Refs.~\cite{Uesaka:2010mm,Korsheninnikov:1997mm}.
\label{fig3}
}
\end{center}
\end{figure}

\begin{figure}[htbp]
\begin{center}
\includegraphics[scale=.75]{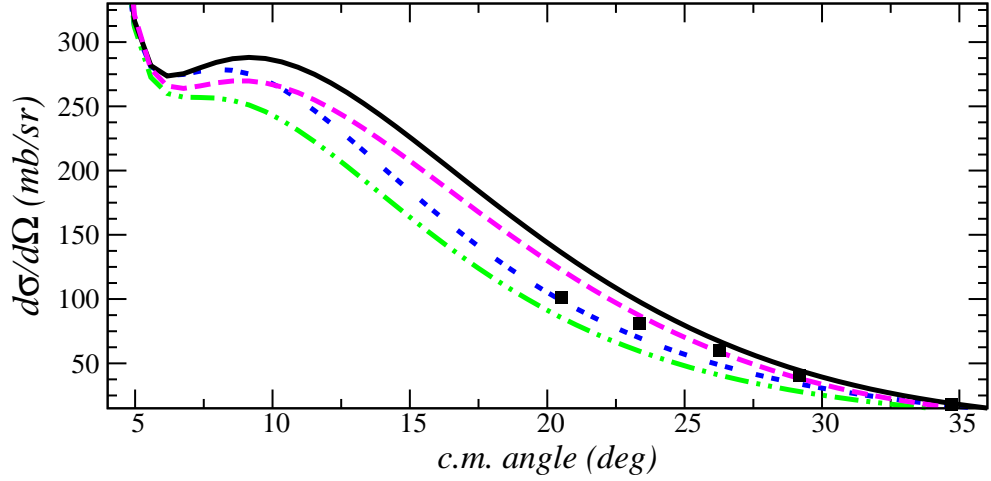}
\caption{(color online) Same as Fig.~3 except that the angular distribution
of the differential cross section ($\frac{d\sigma}{d\Omega}$) 
is plotted with a linear scale. The data are from Ref.~\cite{Korsheninnikov:1997mm}.
\label{fig3b}
}
\end{center}
\end{figure}

Let us first concentrate on the scattering of $^6$He at 71~MeV/nucleon.
In Fig.~\ref{fig3} we show the differential cross section (upper panel (a)) 
and the analyzing
power (lower panel (b)) for elastic scattering of $^6$He at 71~MeV/nucleon. For more detail
we show the differential cross section in a linear scale in Fig.~\ref{fig3b}. The solid
(black) line
represents the calculation with a single-particle optical potential as outlined in Section
II.B based  on the COSMA density.  On the log-scale of Fig.~\ref{fig3} the differential
cross section looks reasonably well described. However, the linear scale of Fig.~\ref{fig3b}
reveals that a for the small angles, the single-particle optical potential over predicts the
differential cross section. The analyzing power stays positive for all angles,  similar to
the predictions in Ref.~\cite{Weppner:2000fi}. 
Our calculation based on the cluster model using
again the COSMA density is represented by the short-dashed (blue) line. The differential
cross section is not very sensitive to the explicit cluster calculation with the exception
of the forward angles, where the cluster calculation gives a lower cross section in better
agreement with the data. The analyzing power, however, does not show any improvement though
the minimum is slightly shifted to smaller angles, it stays positive, whereas the data
indicate a negative sign. The folding for the optical potentials with the COSMA density are
carried out with the free NN t-matrix as input. 
As has been shown in Ref.~\cite{Chinn:1994xz} for a
variety of heavier nuclei, at energies lower than $\sim$150~MeV projectile energy, the
free NN t-matrix experiences a modification due to the nuclear medium, which can be treated
as an additional force represented by a mean field acting on the two active nucleons during
the scattering process~\cite{Chinn:1993zz}. The dash-double-dotted (green) curves in
Figs.~\ref{fig3} and \ref{fig3b} 
include a modification of the free NN t-matrix
through an HFB mean field for the alpha cluster only. One advantage to this cluster
paradigm is that the calculation can utilize the influence of a mean field 
on the free NN interaction where it is most appropriate, i.e. 
for the strongly bound $\alpha$-core.
The results of this calculation produces an analyzing power, which turns negative at 
60$^o$ and captures the shape of the last two measured angles. 
However, it still over predicts the measured analyzing power at the smaller angles. 
The overall shape of the differential cross section is not modified. 
Only for the very forward angles the cross section is slightly
lowered compared to the cluster calculation with the COSMA density as can be seen in
Fig.~\ref{fig3b}. 

It is further instructive to investigate the importance of the correlation density
$\rho_{corr} ({\vvm {\cal P}_j}_c)$ in
Eq.~(\ref{eq:c2}) for the cluster optical potential. This can be done
by realizing that setting $\rho_{corr} ({\vvm {\cal P}_j}_c) =1$ omits
the correlations.
In
Figs.~\ref{fig3} and \ref{fig3b} the dash-double-dotted (green) curve represents the 
calculation based on the cluster formulation whereas the short dashed (pink) line 
represent the same calculation 
with the correlation density set to 1.
The effect on the analyzing power, Fig.~\ref{fig3}, is relatively small. 
However, the differential cross section for small angles, Fig.~\ref{fig3b}, shows
visible sensitivity. Indeed, one can conclude, that the lowering of the differential
cross section for the forward angles is dominated 
by the influence of the correlation density.

\begin{figure}[htbp]
\begin{center}
\includegraphics[scale=.75]{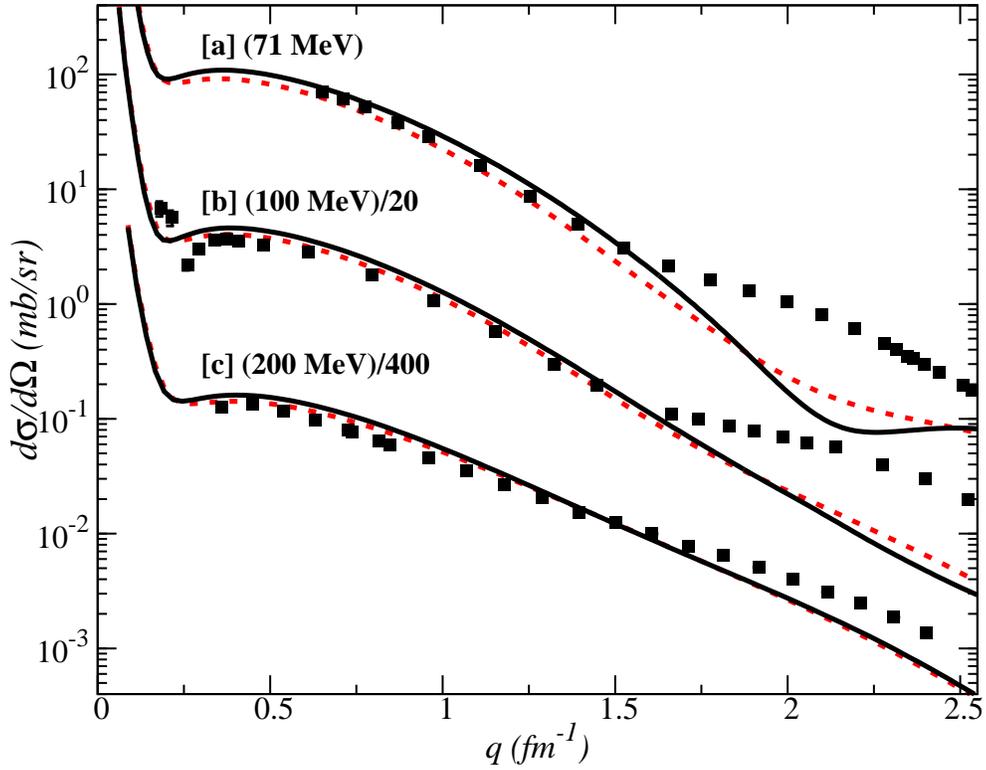}
\caption{(color online) 
The angular distribution of the differential cross
  section ($\frac{d\sigma}{d\Omega}$) for
  elastic proton scattering off $^{4}$He at projectile energies 71~MeV, 100~MeV, and 200~MeV
as function of the momentum transfer.
The calculations are performed with optical potential
  based on the Nijmegen I potential~\cite{Stoks:1993tb} for the NN interaction.
For the $\alpha$-core an HFB density according to Refs.~\cite{HFB,Gogny} is employed. 
The solid lines (black) show the calculations based on the free NN t-matrix, while the
dotted (red) lines are based on calculations modifying the NN t-matrix with a mean-field
consistent with the HFB $\alpha$-core. The data are from 
Refs.~\cite{Burzynski:1989zz,Wesick:1986bt,Comparat:1975bm,Moss:1979aw}.
\label{fig4}
}
\end{center}
\end{figure}

\begin{figure}[htbp]
\begin{center}
\includegraphics[scale=.75]{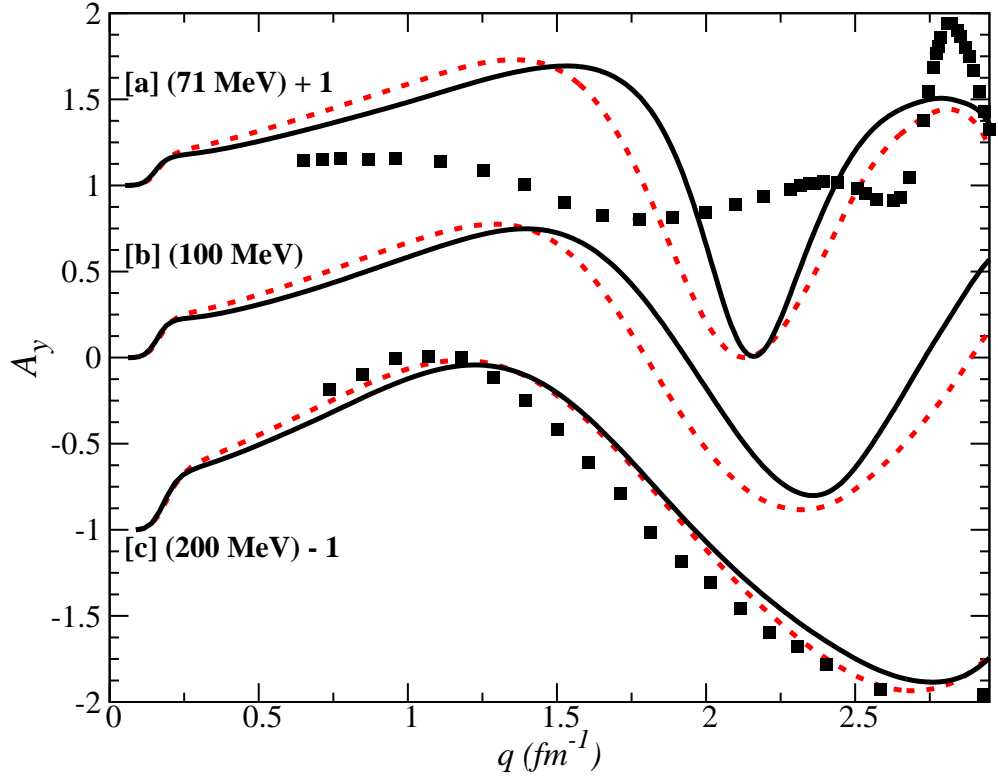}
\caption{(color online) The angular distribution of the analyzing power for
 elastic proton scattering off $^{4}$He at projectile energies 71~MeV, 100~MeV, and 200~MeV
as function of the momentum transfer. The meaning of the curves is the same as in
Fig.~5. The data are from 
Refs.~\cite{Burzynski:1989zz,Wesick:1986bt,Comparat:1975bm,Moss:1979aw}.
\label{fig5}
}
\end{center}
\end{figure}

The upgrade of the RIKEN facility will in principle 
allow to measure the angular distribution
of the analyzing power for the elastic scattering of $^6$He at 
somewhat higher energies. Thus
we want to investigate the predictions for the elastic scattering observables as function of
the projectile energy using a cluster ansatz for the optical potential for 
$^6$He.  
As test energies we choose 100~MeV and
200~MeV/nucleon. First, we show in Figs.~\ref{fig4} 
and \ref{fig5} the angular distributions of
the differential cross section and the analyzing power for proton scattering off $^4$He as
function of the momentum transfer. 
The calculations for the projectile energy of 200~MeV shows
that here an optical potential description of the scattering process is quite good up to
2.5~fm$^{-1}$ for both, differential cross section and analyzing power. 
In  Figs.~\ref{fig4} and \ref{fig5} two calculations are shown: 
For the solid line a folding calculation
of the optical potential has been carried out using the free NN t-matrix, whereas for the
dashed line a NN t-matrix modified by a HFB mean field has been used. 
At 200~MeV a folding with
the free NN t-matrix is adequate. At the two lower energies, a single scattering optical
potential describes the differential cross section only up to a momentum transfer of
 about 1.75~fm$^{-1}$. For higher
momentum transfers multiple scattering contributions are 
expected to become important. However,
only for a deuteron target multiple scattering contributions have been investigated
systematically~\cite{Liu:2004tv,Elster:2008yt} over a wide range of projectile energies.
Thus, we can only speculate about the increasing importance of multiple scattering
contributions at higher momentum transfer.
The analyzing power at 71~MeV projectile energy is not well described for small momentum
transfers. This is very likely the reason for our over-prediction of the analyzing power of
$^6$He for small momentum transfer at this energy~\cite{Sakaguchi:2011rp}.    

\begin{figure}[htbp]
\begin{center}
\includegraphics[scale=.75]{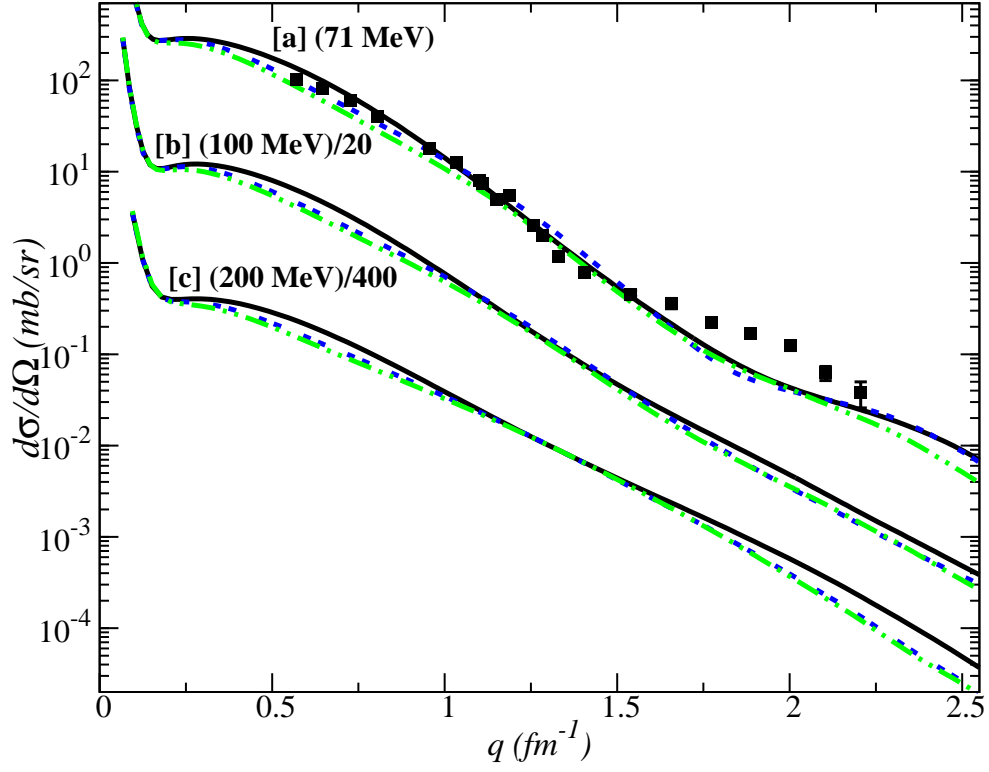}
\caption{(color online)
The angular distribution of the differential cross
  section ($\frac{d\sigma}{d\Omega}$) for elastic scattering of $^{6}$He at projectile energies
 71~MeV,  100~MeV, and 200~MeV/nucleon as function of the momentum transfer.
The calculations are performed with  optical potential
  based on the Nijmegen I potential \cite{Stoks:1993tb} for the NN interaction. 
For the solid (black) line the COSMA~\cite{Zhukov:1994zz} has been used as single-particle
density. The short-dashed (blue) line incorporated the cluster structure into the optical
potential using the COSMA density for all clusters. For the dash-double-dot (green) line the
free NN t-matrix has been modified with  the HFB mean field.
The data are taken from Ref.~\cite{Uesaka:2010mm,Korsheninnikov:1997mm}.
\label{fig6}
}
\end{center}
\end{figure}

\begin{figure}[htbp]
\begin{center}
\includegraphics[scale=.75]{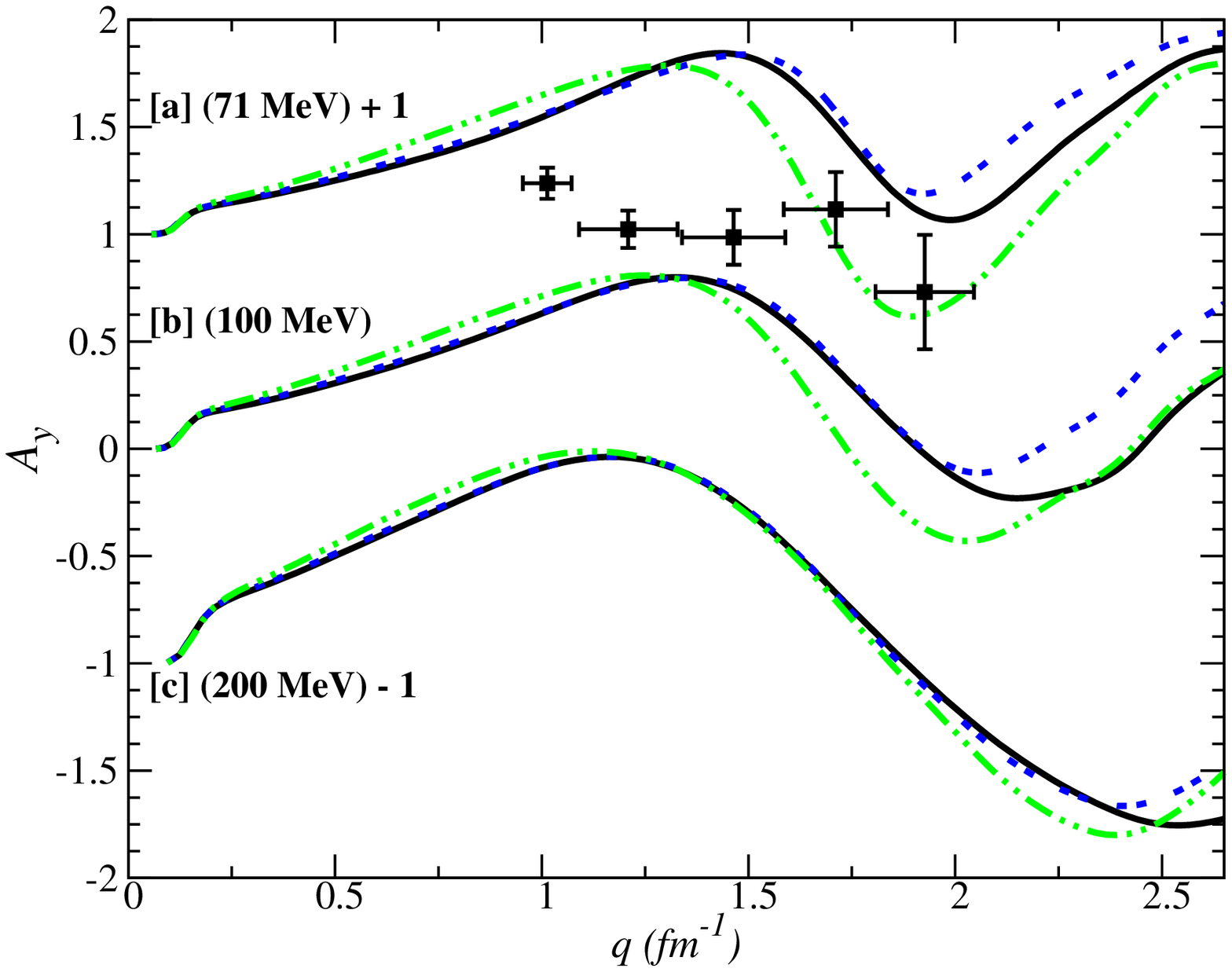}
\caption{(color online) 
The angular distribution of the the analyzing power ($A_y$) for
  elastic scattering $^{6}$He at projectile energies
 71~MeV,  100~MeV, and 200~MeV/nucleon as function of the momentum transfer. The meaning of
the lines is the same as in Fig.~7.
The data are taken from Ref.~\cite{Uesaka:2010mm, Korsheninnikov:1997mm}.
\label{fig7}
}
\end{center}
\end{figure}

\begin{figure}[htbp]
\begin{center}
\includegraphics[scale=.75]{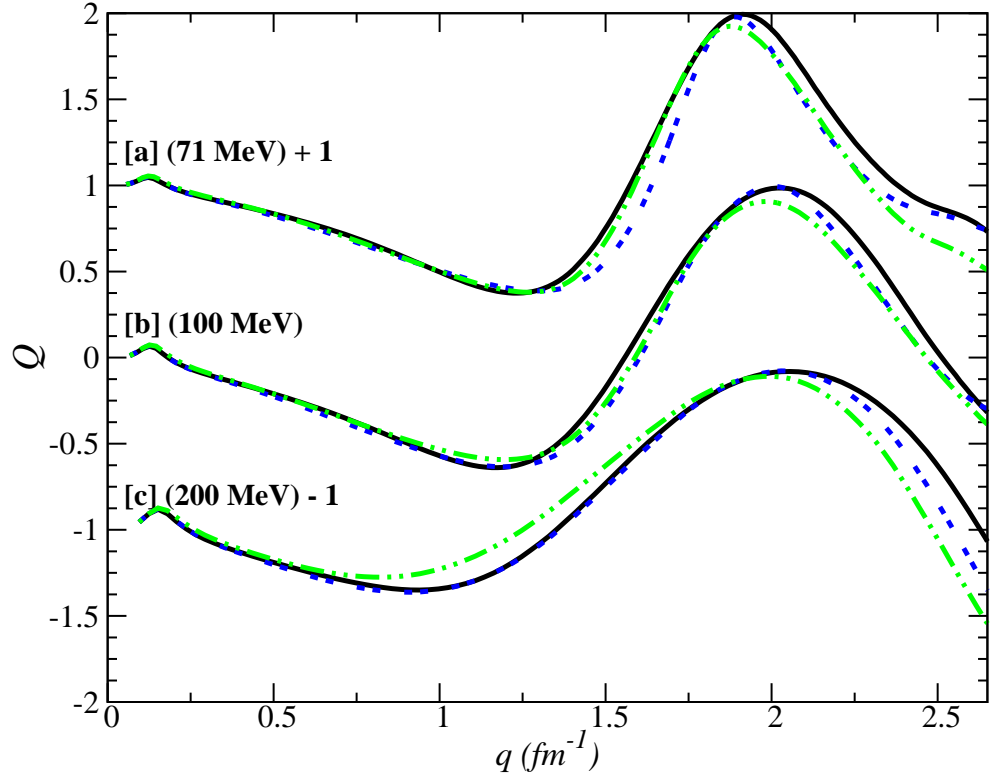}
\caption{(color online) 
The angular distribution of the the spin rotation function $Q$ for
  elastic scattering $^{6}$He at projectile energies
 71~MeV,  100~MeV, and 200~MeV/nucleon as function of the momentum transfer. The meaning of
the lines is the same as in Fig.~7. 
\label{fig8}
}
\end{center}
\end{figure}

In Figs.~\ref{fig6} and \ref{fig7} we present predictions for elastic scattering of $^6$He
at 100~MeV and 200~MeV/nucleon. The solid lines represent the calculations with a
single-particle optical potential, whereas the short-dashed and dash-double dotted lines are
based on the cluster formulation 
discussed in Section~II.C. The short-dashed line uses a free NN
t-matrix for every piece of the optical potential, whereas the dash-double-dotted line
incorporates a NN t-matrix modified by an HFB mean field potential for the $\alpha$-core.
Similar to the $^4$He calculations, at 200~MeV/nucleon the modification of the NN t-matrix
becomes irrelevant. As discussed for the 71~MeV/nucleon calculations, 
the cluster model lowers
the predictions for the differential cross section 
for small momentum transfers (angles). This
feature seems independent of the employed projectile energy. For the analyzing power 
at small momentum transfer the contribution of the valence neutrons is not large enough 
to change the contribution of the 
$\alpha$-core part of the optical potential. 
For the 100~MeV calculation the modification of the free NN t-matrix through
the nuclear medium is still quite pronounced at $\sim$2~fm$^{-1}$ momentum transfer, whereas
the 200~MeV/nucleon calculations exhibit very little difference between each other.
 In Fig.~\ref{fig8} we show the spin-rotation function $Q$ for elastic scattering 
of $^6$He at the same energies. Obviously, this observable has no chance of being
measured. However, since it is an independent observable, we consider 
it instructive to study if it exhibits 
differences similar to the ones in the analyzing power 
between a cluster paradigm for the optical potential and a single-particle optical
potential. It is interesting to note, that $Q$ is very insensitive to any of the
modifications introduced, even at 71~MeV/nucleon.
 
\begin{figure}[htbp]
\begin{center}
\includegraphics[scale=.75]{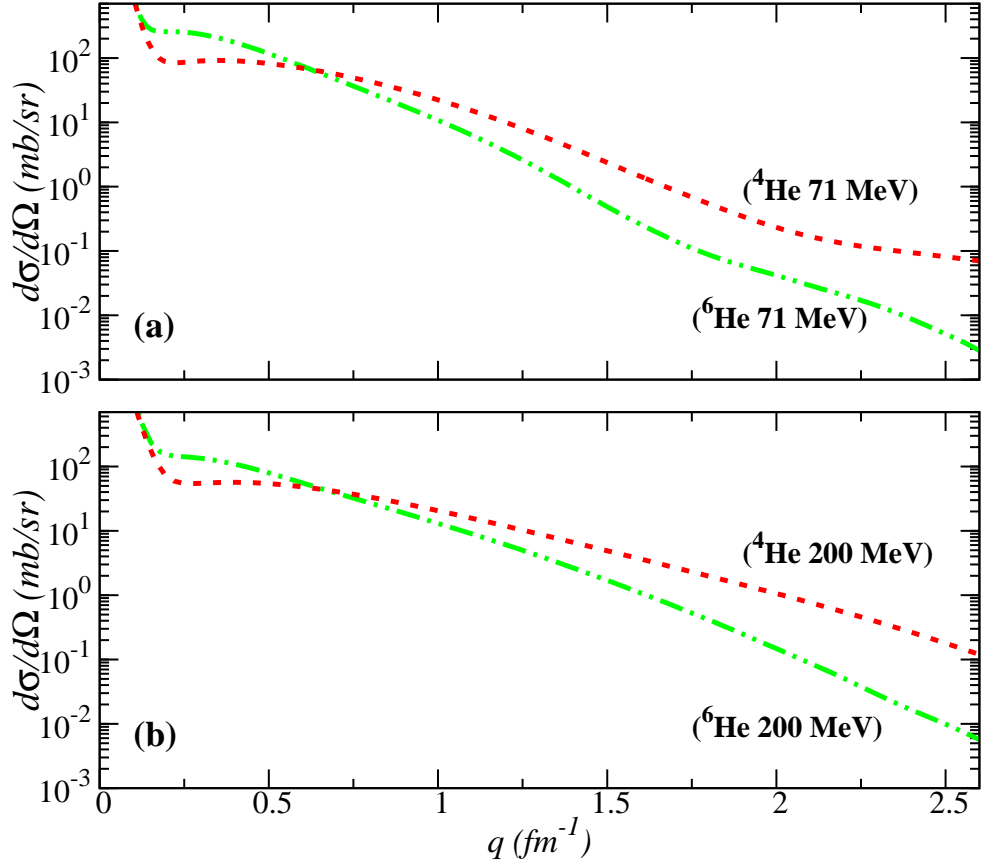}
\caption{(color online) 
The angular distribution of the differential cross
  section, $\frac{d\sigma}{d\Omega}$, for elastic proton scattering off
$^{4}$He
(short-dashed red) and $^{6}$He (dash-double-dot green)  at projectile
energies
 71~MeV (upper panel (a)) and 200~MeV/nucleon (lower panel (b)) 
as function of the momentum transfer.
Both calculations are performed with  optical potential
  based on the Nijmegen I potential \cite{Stoks:1993tb} for the NN
interaction, which
for $^{4}$He (in the $^{6}$He case the $^{4}$He-core) is modified by the
HFB mean field.
For the $^{6}$He calculations the single particle nucleons are described
by the COSMA density, and the cluster ansatz is used.
For the $^{4}$He calculations the HFB density \cite{HFB,Gogny} is
employed.
\label{fig9}
}
\end{center}
\end{figure}

\begin{figure}[htbp]
\begin{center}
\includegraphics[scale=.75]{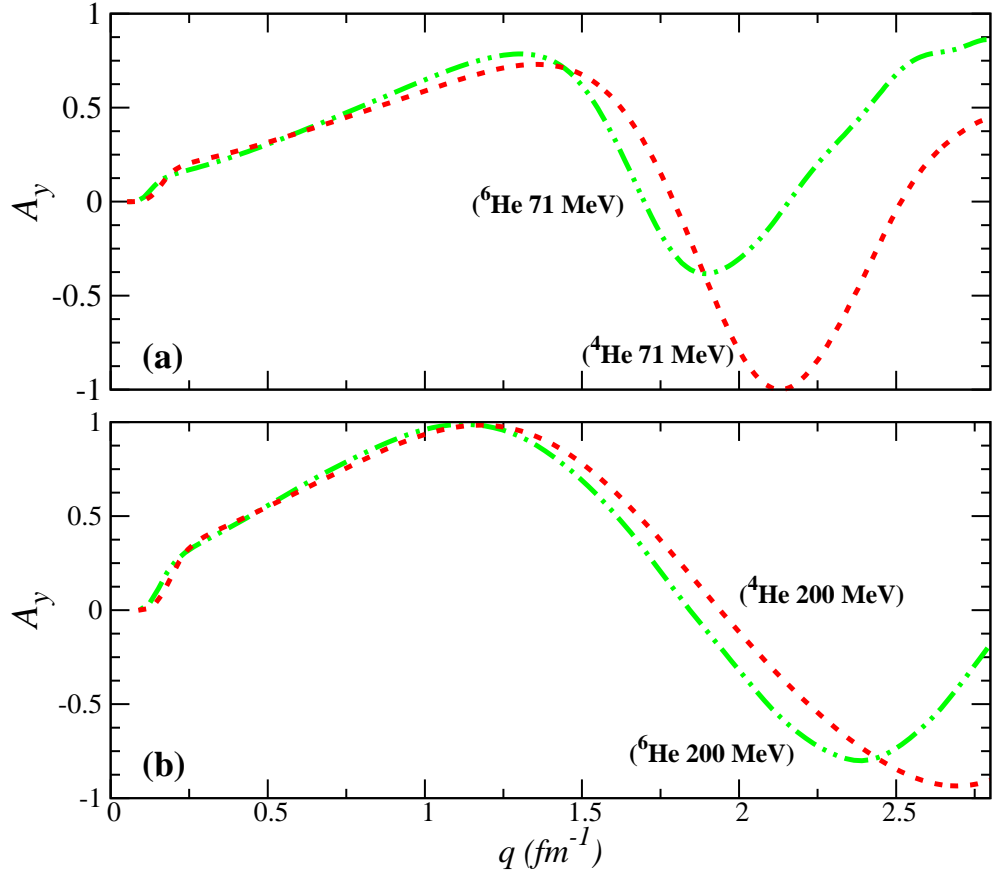}
\caption{(color online) 
The angular distribution of the analyzing power for elastic proton scattering off $^{4}$He 
(short-dashed red) and $^{6}$He (dash-double-dot green)  at projectile energies
 71~MeV (upper panel (a)) and 200~MeV/nucleon (lower panel (b))
 as function of the momentum transfer. The meaning of the curves
is the same as in Fig.~\ref{fig9}.
\label{fig10}
}
\end{center}
\end{figure}

In order to investigate if the difference of the observables 
for elastic scattering of $^4$He and
$^6$He changes with increasing energy, we plot in Figs.~\ref{fig9} and \ref{fig10} the
differential cross sections and analyzing powers for the two nuclei together at 71~MeV and
200~MeV as function of the momentum transfer. 
One could speculate that when employing a cluster
model for the $^6$He nucleus, at higher energies the piece of the optical potential
due to the $\alpha$-core might dominate the observables (possibly the analyzing power).
However, this does not seem to be the case, 
the behavior of the differential cross section is
very similar at the two energies. 
Even the shift of the minima in the analyzing power between
the two nuclei  is the same for both energies, namely roughly 0.25~fm$^{-1}$. 
Obviously, experimental information will have to
decide if a cluster ansatz as presented here captures a major part 
of the underlying physics.

\section{Summary and Outlook}
In this work we introduced a cluster formulation for the optical potential for calculating 
scattering observables for elastic scattering of three-body halo nuclei. We first reviewed
a traditional single-particle full-folding optical potential in 
first order in the Watson multiple
scattering expansion,
and then showed how one can naturally extend this optical potential to
introduce the cluster structure of a halo nucleus. Here we concentrate on the $^6$He
nucleus. However, the formulation we introduced can be further extended to four or five-body
clusters, e.g. to $^8$He. 
For our calculations we used the density matrix of the
three-body cluster orbital shell model approximation
(COSMA) introduced in Refs.~\cite{Zhukov:1994zz,Zhukov:1991} for the $^6$He nucleus. 
This density matrix is based on single harmonic
oscillator wave functions for the s- and p-shell of $^6$He and allows a straightforward
calculation of the required correlation densities needed for the optical potential. The
resulting folding optical potential contains  a six dimensional integration over internal
vector momenta, which is calculated via Monte Carlo integration.

We calculated the angular distribution of the 
differential cross section and the analyzing power
at 71~MeV, 100~MeV, and 200~MeV/nucleon and compared our results
with experimental data at
71~MeV/nucleon~\cite{Uesaka:2010mm,Korsheninnikov:1997mm}. We find that the cluster model
lowers the cross section for the small angles and brings it closer to the data. Though we do
not describe the very small analyzing power at the small angles, we find that the
cluster formulation together with a `hybrid' ansatz, in which the optical potential for
the alpha-core is calculated with a NN t-matrix modified by a mean field of the alpha 
is able to 
produce a negative analyzing power at larger angles as suggested by
the data.  
Our predictions for the higher energies indicate that the lowering of the
differential cross section for small momentum transfers (angles) using a cluster paradigm
remains visible. The cluster ansatz for the optical potential
 continues to predict a negative analyzing power at larger momentum
transfer. Eventually experimental information should be become available to see if these
predictions capture the bulk of the physics of the reaction at higher energies,
or if there are additional theoretical pieces necessary to understand this reaction,
preferably as function of scattering energy.



\appendix

\section{First Order Full-Folding Optical Potential}
\label{appendix0}

In this appendix we will give explicit steps for arriving at
the `traditional'  first order Watson optical potential of
our calculations. The complete derivation is given in 
Ref.~\cite{Elster:1997as}, however for the convenience of the
reader we want to give a shorter summary here.

Using the $\delta$-function of Eq.~(\ref{eq:2.8.1a}), which
determines the conservation of the absolute momentum for the 
center of mass  (c.m.) frame of the nucleus and 
the relative momenta of the individual nucleons in the target,
we obtain for the optical potential of Eq.~(\ref{eq:2.8.1})
\begin{eqnarray}
\langle\hat\tau_{01}\rangle & \equiv & \langle {\bf k'}|\langle \phi_A|{\hat
\tau}_{0i}{(\cal{E}})| \phi_A\rangle |{\bf k}\rangle \cr
&=& \int\prod_{j=1}^Ad\vvm k_j' \;\int\prod_{l=1}^A d \vvm k_l
\;\langle \phi_A |\vvm\zeta_1'\vvm\zeta_2'\vvm\zeta_3'\vvm\zeta_4'...
\vvm\zeta_{A-1}'
\rangle \delta  (\vvm p' - \vvm p_0')\;
\langle\vvm k' \vvm k_1' |\hat\tau_{01}({\cal{E}})|
\vvm k\vvm k_1\rangle  \cr
& & \prod_{j=2}^A \delta(\vvm k_j'- \vvm k_j) \delta  (\vvm p - \vvm p_0)
\;\langle \vvm\zeta_1\vvm\zeta_2\vvm\zeta_3\vvm\zeta_4...\vvm\zeta_{A-1}
|\phi_A\rangle,
\label{eq:2.8.4}
\end{eqnarray}
where $\vvm p=\sum_{i=1}^A \vvm k_i, \vvm p' = \sum_{i=1}^A \vvm k'_i$.
Without losing generality, we consider for now nucleon `$1$' being the active nucleon in
the target.
The additional delta functions arises because the target nucleons
are independent of $\hat{\tau}_{01}$. 
A Galilean invariant choice of internal variables are 
the Jacobi coordinates,
\begin{eqnarray}
\vvm k_{1}&=& {\bf\zeta_1} + \frac{\vvm p}{A} \nonumber \\
\vvm k_{2}&=& {\bf\zeta_2} - \frac{{\bf\zeta_1}}{A-1} + 
\frac{\vvm p}{A} \nonumber \\ 
\vvm k_{3}&=& \vvm {\bf\zeta_3} - \frac{{\bf\zeta_1}}{A-1} - 
\frac{\bf \zeta_2}{A-2} + \frac{\vvm p}{A} \nonumber \\ 
&&\vdots \nonumber \\
\vvm k_{A}&=& {\bf \zeta_A} - \sum_{j=1}^{A-1}
 \frac{\bf \zeta_j}{A-j} + \frac{\vvm p}{A} .\label{eq:2.8.5}
\end{eqnarray}
In first order, which is the concern of this work, we only need the momentum
of the struck nucleon,
namely $\vvm k_{1}={\bf \zeta_1} + \frac{\vvm p}{A}$.
Changing  integration variables from absolute to 
relative momenta 
in Eq.~(\ref{eq:2.8.4}) results in
\begin{eqnarray} 
\langle\hat\tau_{01}\rangle &=&
\;\int\prod_{j=1}^{A-1}d{\bf \zeta_j'} d\vvm p'
\;\int\prod_{l=1}^{A-1}d {\bf \zeta_l} d\vvm p
\;\langle \phi_A |{\bf\zeta_1'\zeta_2'\zeta_3'\zeta_4'...
\vvm\zeta_{A-1}'}
\rangle \delta  (\vvm p' - \vvm p'_0) \nonumber \\
&&\langle\vvm k',{\bf \zeta_1'}+\frac{\vvm p'}{A} 
\vert \hat\tau_{01}({\cal{E}}) \vert
\vvm k,{\bf \zeta_1}+\frac{\vvm p}{A}\rangle 
\prod_{j=2}^{A-1} \delta({\bf \zeta_j'}- {\bf \zeta_j})
\delta\left(\frac{A-1}{A}\vvm p'-{\bf\zeta_1'} 
- \frac{A-1}{A}\vvm p+{\bf\zeta_1}\right)
\nonumber \\ &&\delta  (\vvm p - \vvm p_0) 
\langle {\bf\zeta_1\zeta_2\zeta_3\zeta_4...\zeta_{A-1} }
|\phi_A\rangle.
\label{eq:2.8.6}
\end{eqnarray}
Eq.~(\ref{eq:2.8.6}) indicates that one only needs the
single particle density matrices $\rho({\bf\zeta_1',\zeta_1})$, describing
the dependence of one particle's  
relative motion to the remaining (A-1)-core, and given in 
Eq.~(\ref{eq:2.8.1b}).
Inserting $\rho({\bf \zeta_1'},{\bf\zeta_1})$ 
into Eq.~({\ref{eq:2.8.6}}) and evaluating 
$\delta  (\vvm p - \vvm p_0)$ and $\delta(\vvm p'-\vvm p'_0)$ leads to
\begin{eqnarray}
\langle\hat\tau_{01}\rangle &=&
\;\int d{\bf\zeta_1'} \;\int d{\bf\zeta_1}
\langle\vvm k'\; {\bf\zeta_1'}+\frac{\vvm p_0'}{A} \left|
\hat\tau_{01}({\cal{E}})\right|
\vvm k\; {\bf\zeta_1}+\frac{\vvm p_0}{A}\rangle\;
\rho({\bf\zeta_1',\vvm \zeta_1}) \nonumber \\
&&\; \delta\left(\frac{A-1}{A}\vvm p_0'-{\bf\zeta_1'} - \frac{A-1}{A}\vvm p_0
+{\bf\zeta_1}\right).
\label{eq:2.8.7}
\end{eqnarray}
Inserting the NN ${\hat \tau}_{01}$-matrix of Eq.~({eq:2.8.1c})
and taking advantage of its conservation of the c.m. momentum 
Eq.~(\ref{eq:2.8.7}) becomes
\begin{eqnarray}
\langle\hat\tau_{01}\rangle &=&
\;\int d{\bf\zeta_1'} \;\int d {\bf\zeta_1}\;\delta(\vvm k' + \vvm p_0' - \vvm k - 
\vvm p_0) \nonumber \\
&&\left\langle\frac{1}{2}(\vvm k'-{\bf\zeta_1'}
-\frac{\vvm p_0'}{A}) |\hat\tau_{01}({\cal{E}})|
\frac{1}{2}(\vvm k- {\bf\zeta_1}-\frac{\vvm p_0}{A})\right\rangle \;\;
\rho({\bf\zeta_1'},{\bf\zeta_1}) \nonumber \\
&&\delta\left(\frac{A-1}{A}\vvm p_0'-{\bf\zeta_1'} -
\frac{A-1}{A}\vvm p_0 +{\bf\zeta_1}\right). 
\label{eq:2.8.8}
\end{eqnarray}
The first delta  function  describes the overall momentum conservation,
which can be used  to reduce the integral of
Eq.~({\ref{eq:2.8.8}}) to three dimensions, i.e. 
the integral given in Eq.~(\ref{eq:2.8.8b}).
In our practical calculations we use the variables 
$\vvm q$, $\vvm K$, and $\vvm P$, which are given in Eq.~(\ref{eq:2.8.8a}).
The inverse relations are given by
\begin{eqnarray}
\vvm k &=& \vvm K - \frac{1}{2} \vvm q \nonumber \\
\vvm k'& =& \vvm K + \frac{1}{2} \vvm q \nonumber \\ 
\vvm \zeta_1& =& \vvm P+\frac{A-1}{2A}\vvm q \nonumber \\
\vvm \zeta_1'& =& \vvm P-\frac{A-1}{2A}\vvm q. \label{eq:2.8.8bb}
\end{eqnarray}
Substituting those  variables into Eq.~(\ref{eq:2.8.8b}) leads to
\begin{eqnarray}
\langle\hat\tau_{01}\rangle &=&
\left\langle\frac{1}{2}\left(\vvm K
-\vvm P+\frac{2A-1}{2A}\vvm q -\frac{\vvm p_0'}{A}\right) |\hat\tau_{01}
({\cal{E}})| \frac{1}{2}\left(\vvm K
- \vvm P-\frac{2A-1}{2A}\vvm q -\frac{\vvm p_0}{A}\right)\right\rangle\; \nonumber \\
&&\rho\left(\vvm P-\frac{A-1}{2A}\vvm q
,\vvm P+\frac{A-1}{2A}\vvm q\right).
\label{eq:2.8.9a}
\end{eqnarray}
Here we dropped the overall momentum conserving  delta function,
which is carried out  when evaluating the cross section.
Since the  preceding derivation based on the projectile particle `0' and  active
target particle `1' is general, and thus can be repeated 
for all $N$ target neutrons and $Z$ target protons,
one obtains as final expression for the `full-folding' optical potential
\begin{eqnarray}
\lefteqn{U_{el}\left(\vvm K + \frac{1}{2} \vvm q,
\vvm K - \frac{1}{2} \vvm q\right) = } \cr 
&&\sum_{i=N,P} \;\int d \vvm P  
\left\langle\frac{1}{2}\left(\vvm K 
-\vvm P+\frac{2A-1}{2A}\vvm q -\frac{\vvm p_0'}{A}\right) |\hat\tau_{0i}
({\cal{E}})| \frac{1}{2}\left(\vvm K 
- \vvm P-\frac{2A-1}{2A}\vvm q -\frac{\vvm p_0}{A}\right)\right\rangle\; \nonumber \\
&&\rho_i\left(\vvm P-\frac{A-1}{2A}\vvm q
,\vvm P+\frac{A-1}{2A}\vvm q\right).
\label{eq:2.8.9}
\end{eqnarray}
From Eq.~(\ref{eq:2.8.9}) we can read off the momenta of the NN t-matrix as
\begin{eqnarray}
\vvm k_{NN}&=&\frac{1}{2}\left(\vvm K  
- \vvm P-\frac{2A-1}{2A}\vvm q -\frac{\vvm p_0}{A}\right) \nonumber \\
\vvm k_{NN}'&=&\frac{1}{2}\left(\vvm K  
-\vvm P+\frac{2A-1}{2A}\vvm q -\frac{\vvm p_0'}{A}\right) \label{eq:2.9.99}.
\end{eqnarray}
For numerical calculations we prefer 
\begin{eqnarray}
\vvm q_{NN}&=&\vvm k_{NN}' - \vvm k_{NN}=
\vvm k' -\vvm k = \vvm q \nonumber \\
\vvm K_{NN}&=&\frac{1}{2}\left(\vvm k_{NN}' + \vvm k_{NN}\right)=
\frac{1}{2}\left(\vvm K - \vvm P -\frac{\vvm p_0+\vvm p_0'}{2A}\right).
\end{eqnarray}
Since \vv q is a momentum transfer, it is invariant under 
Galilean transformations, i.e. $\vvm q_{NN} = \vvm q$.
Rewriting the optical potential in terms of {\bf q} and {\bf K} gives 
\begin{eqnarray}
\lefteqn{U_{el}(\vvm q, \vvm K) =} \cr
&&\sum_{i=N,P} \int d \vvm P \hat\tau_{0i}\left(\vvm q ,
\frac{1}{2}\left(\vvm K - \vvm P -\frac{\vvm p_0+\vvm p_0'}{2A}\right),{\cal{E}}\right)
\;\rho_i\left(\vvm P-\frac{A-1}{2A}\vvm q
,\vvm P+\frac{A-1}{2A}\vvm q \right). 
\label{eq:2.8.10} 
\end{eqnarray} 
Rewriting Eq.~(\ref{eq:2.8.10}) in variables of the NA c.m. system requires
that $\vvm k+\vvm p_0 = \vvm k'+\vvm p_0'=0$.
Thus, the dependence of $\vvm p_0$ drops out, and one obtains for the `full-folding'
single-particle optical potential of Eq.~(\ref{eq:2.8.11})
\begin{eqnarray}
\lefteqn{U_{el}(\vvm q, \vvm K) = } \cr
&&\sum_{i=n,p}
\int d \vvm P \hat\tau_{0i}\left(\vvm q ,
\frac{1}{2}\left(\frac{A+1}{A}\vvm K - \vvm P\right),{\cal{E}}\right)
\;\rho_i\left(\vvm P-\frac{A-1}{2A}\vvm q
,\vvm P+\frac{A-1}{2A}\vvm q \right). 
\end{eqnarray}


\section{Transforming the optical potential}
\label{appendixa}

We first want to calculate the optical potential for a nucleon
scattering off a cluster `$i$' in the c.m. frame  of
the projectile and the cluster `$i$' (e.g. cluster `$i$'
could be the $\alpha$ particle within the $^6$He nucleus). Let us
use the subscript $C_i$  to  denote that frame, whereas quantities without
subscripts shall
be interpreted as given in the $A+1$ c.m. frame. The overall
conservation of momentum for the c.m. cluster frame assumes
\begin{equation}
\vvm k_{C_i} + \vvm {k_{1}}_{C_i} = {{\vvm k^\prime}_{C_i}} 
+ \vvm {{k^\prime_{1}}_{C_i}} \equiv 0,
\end{equation}
where $\vvm k_{C_i}$ is the projectile and $\vvm {k_{1}}_{C_i}$ 
is a typical target nucleon 
inside the nucleus, within the $i$th cluster.

Starting with the definition of the density given in Eq.~(\ref{eq:2.8.5})
in the $A+1$ frame,
\begin{equation}
{\mathbf\zeta_1}= \vvm k_{1} - \frac{\vvm p}{A} \nonumber ,
\end{equation}
this becomes in a specific cluster frame 
\begin{equation}
{\mathbf\zeta_1}= \vvm {k_{1}}_{C_i} - \frac{\vvm p_{C_i}}{A}.\label{ap:8}
\end{equation}
The momentum ${\mathbf\zeta_1}$ does not carry a cluster subscript since
it is defined in the traditional intrinsic frame of the single
particle density.
Eq.~(\ref{ap:8}) defines how this intrinsic variable
is related via a Galilean transformation to the cluster frame.
Using the notation of Eq.~(\ref{eq:c1}), this can be broken up 
into the active particle and the spectator
\begin{equation}
{\mathbf\zeta_1}= \vvm {k_{1}}_{C_i} - \frac{\vvm {p_i}_{C_i} 
+  \vvm {{p_s}_i}_{C_i}}{A}. \label{ap:1}
\end{equation}
In this center of mass frame one has $\vvm {k}_{C_i} = - \vvm {p_i}_{C_i}$,
where $\vvm {k}_{C_i}$ is the momentum of the projectile in the cluster
frame. Then
using Eq.~(\ref{ap:1}), this result together with understanding that the 
spectator momentum does not change during the collision,
we can calculate the difference
\begin{equation}
{\bf\zeta_1}- {{\bf\zeta_1}}^\prime= 
(\vvm {k_{1}}_{C_i}- {\vvm {k_{1}^\prime}_{C_i}}) 
- (\frac{{\vvm {k^\prime}_{C_i}}
-  \vvm {k}_{C_i}}{A})
= \frac{A-1}{A} \vvm q,
\end{equation}
remembering that $\vvm q$, the relative momentum transfer,
is invariant in all frames.
If we allow the same definition for the average
momentum of the target nucleon,
Eq.~(\ref{eq:2.8.8a})
\begin{equation}
\vvm P \equiv \frac{1}{2}({\bf{\zeta_1}}^\prime
+{\bf\zeta_1}),
\end{equation}
then the inverse equations of Eq.~(\ref{eq:2.8.8bb}),
\begin{eqnarray}
&& {\bf\zeta_1} = \vvm P+\frac{A-1}{2A}\vvm q \nonumber \\
&& {\bf{\zeta_1}}^\prime = \vvm P-\frac{A-1}{2A}\vvm q \label{ap:2}
\end{eqnarray}
have the same form. This makes sense, since $\vvm q$ is
invariant and $\vvm P$ is written in the intrinsic frame of the nucleus.

The momentum arguments of the $\hat\tau_{0i}$-matrix 
 are defined  in Eq.~(\ref{eq:2.9.99}) for the $A+1$ frame. However, we 
also need to redefine them in the cluster frame $C_i$. Rewriting 
Eq.~(\ref{eq:2.9.99}) and using Eqs.~(\ref{ap:1}) and (\ref{ap:2}), we
can write the relative momentum between the 
projectile, $\vvm  k_{C_i}$ and the struck
nucleon, $\vvm {k_{1}}_{C_i}$ as 
\begin{equation} 
\vvm k_{C_i} - \vvm {k_{1}}_{C_i} = \vvm {k_{NN}}_{C_i} =
\vvm k_{C_i} - {\bf\zeta_1} + \frac{\vvm {k}_{C_i}
- \vvm {{p_s}_i}_{C_i}}{A}.
\label{ap:3}
\end{equation}
The primed momentum is similarly given as
\begin{equation} 
\vvm {k^\prime_{C_i}} - \vvm {{k^\prime_{1}}_{C_i}} = 
\vvm {{k^\prime_{NN}}_{C_i}} =  
\vvm {k^\prime_{C_i}} - {{\bf\zeta_1}}^\prime + 
\frac{\vvm {{k^\prime}_{C_i}} - \vvm {{{p^\prime_s}_i}_{C_i}}}{A}.
\label{ap:4}
\end{equation}
The difference between Eq.~(\ref{ap:3}) and Eq.(\ref{ap:4}) gives 
the momentum transfer
$\vvm q$ as expected. We can define the sum 
of the momenta of the $\hat\tau_{0i}$-matrix 
as the sum of Eqs.~(\ref{ap:3}) and (\ref{ap:4}) as
\begin{equation} 
\vvm {K_{NN}}_{C_i} = \frac{\vvm {k_{NN}}_{C_i} 
+ \vvm {{k^\prime_{NN}}_{C_i}}}{2}
= \frac{A+1}{2A}(\vvm {k_{C_i}}+\vvm {k^\prime_{C_i}}) 
-\frac{1}{2}({\bf\zeta_1} + {{\bf\zeta_1}}^\prime) 
- \frac{\vvm {{p_s}_i}_{C_i}}{A}, \label{ap:5}
\end{equation}
remembering that the spectator momentum does not change during the collision. 
In the spirit of Eq.~(\ref{eq:2.8.8a}) we can define the 
sum of the cluster momenta as $\vvm {K}_{C_i} 
= \frac{1}{2}({\vvm {k_{NN}}_{C_i} + \vvm {{k^\prime_{NN}}_{C_i}}})$
and then rewrite Eq.~(\ref{ap:5}) as
\begin{equation} 
\vvm {K_{NN}}_{C_i} = \frac{\vvm {k_{NN}}_{C_i} 
+ \vvm {{k^\prime_{NN}}_{C_i}}}{2}
= \frac{A+1}{A}\vvm {K}_{C_i} - \vvm P 
- \frac{\vvm {{p_s}_i}_{C_i}}{A}.   \label{ap:6}
\end{equation}
Examining Eqs.~(\ref{eq:2.9.99}) through (\ref{eq:2.8.11}), 
we see that the momentum
argument $\vvm {K_{NN}}$ of the $\hat\tau_{0i}$-matrix  has picked up an extra 
term involving the spectator momentum 
in the cluster frame. The rational for this is simple, the struck nucleon is 
acted upon in a specific cluster, 
but the  total density contains both cluster and spectators.

We can rewrite $\vvm {K_{NN}}$ in terms of the 
correlation momentum ${\vvm {\cal P}_j}_i$
by multiplying $\vvm {K_{NN}}_{C_i}$ by the number of nucleons in 
the $i$th cluster, $A_i$,
\begin{equation} 
\vvm A_i{K_{NN}}_{C_i} 
= A_i\frac{A+1}{A}\vvm {K}_{C_i} - A_i\vvm P 
- A_i\frac{\vvm {{p_s}_i}_{C_i}}{A}.\label{ap:6b}
\end{equation}
After some manipulation as well as  using the definition of 
${\vvm {\cal P}_j}_i$ found in Eq.~(\ref{eq:c2}), one finds
\begin{equation} 
\vvm A_i{K_{NN}}_{C_i} 
= (A_i+1)\vvm {K}_{C_i} - A_i\vvm P +{\vvm {\cal P}_j}_i,
\end{equation}
so a cleaner definition of the average momentum can be written as
\begin{equation} 
{K_{NN}}_{C_i} 
= \frac{A_i+1}{A_i}\vvm {K}_{C_i} - \vvm P +\frac{{\vvm {\cal P}_j}_i}{A_i}.
\end{equation}

\noindent
Thus,  for a specific cluster frame we can write the optical potential as
\begin{eqnarray}
\lefteqn{{U_{el}(\vvm q, \vvm K)}_{C_i} = \sum_{t=n,p}
\int d \vvm P \; d{\vvm {\cal P}_j}_i \; \rho_{corr} ({\vvm {\cal P}_j}_i) } \cr
&& \hat\tau_{0t}\left(\vvm q , \frac{1}{2}\left(\frac{A_i+1}{A_i}\vvm K_{C_i} 
- \vvm P+\frac{{\vvm {\cal P}_j}_i}{A_i}\right),
{\cal{E}}\right)
\;\rho_i\left(\vvm P-\frac{A-1}{2A}\vvm q ,\vvm P+\frac{A-1}{2A}\vvm q \right).
\label{ap:7}
\end{eqnarray}
In the case of $^6$He this is the optical potential for a proton on 
the alpha core (or the proton projectile on one of the neutrons). 
We have not worried about the correlation density transformation 
since this is based on a relative 
momentum ${\vvm {\cal P}_j}_i$ and is thus invariant 
during a frame transformation, as are $\vvm P$ and $\vvm q$, the
variables of the traditional single particle density.

In order to be useful, Eq.~(\ref{ap:7}) must be transformed from each individual 
cluster frame back to the nucleon-nucleus frame,  so that it can be summed with
the other clusters which make up the target nucleus.
Scattering observables can then be calculated in the 
c.m. frame of the $A+1$ system following Eq.~(\ref{eq:2.8.11c}).
As it stands, each cluster has its own unique c.m. frame, and thus
they cannot be summed until they are transformed back to 
the unique $A+1$ c.m. frame. 
The only argument of concern, because it is not invariant, 
is the momentum $\vvm K_{C_i}$ 
in the  $\hat\tau_{0i}$ operator.  

Employing conservation of  momentum in both frames, we can define how the 
cluster frame relates to the nucleon-nucleus $A+1$ frame.
Setting relative velocities equivalent in the two different frames leads to
\begin{equation}
A_i {\bf k} - {\bf p_i} = A_i {\bf k}_{C_i} - {\bf {p_i}}_{C_i} = 
(A_i + 1) {\bf k}_{C_i},\label{ap:10}
\end{equation}
where the last equivalence is given because in the c.m. of
the cluster frame ${\bf k}_{C_i} = - {\bf p_i}_{C_i}$.
Another second relation between the two frames is gained by examining the Jacobi 
momentum  of Eq.~(\ref{eq:c1}) in the $A+1$ frame,
\begin{equation}
{\vvm p_j}_i = \frac{1}{A}({A_s}_i \vvm p_i - A_i \vvm {p_s}_i).
\end{equation}
Rearranging gives
\begin{equation}
\vvm p_i = \frac{A_i}{A}\left(\vvm p_i + \vvm {p_s}_i\right)+{\vvm p_j}_i =
-\frac{A_i}{A}{\vvm k}+{\vvm p_j}_i,\label{ap:9}
\end{equation}
where again the last equivalence is found by recognizing that in the
$A+1$ frame the c.m. momentum is defined as 
$\vvm k + \vvm p_i + \vvm {p_s}_i= 0$.
Inserting the result for $\vvm p_i$ from Eq.~(\ref{ap:9}) into 
Eq.~(\ref{ap:10}) we can after some manipulation  compare the momenta between
the two frames:
\begin{equation}
\vvm k = \frac{A}{A+1}\frac{1}{A_i}\left((A_i+1)k_{C_i}+{\vvm p_j}_i\right).
\end{equation}
The same relationship can be developed for the primed momentum
\begin{equation}
\vvm k^\prime = \frac{A}{A+1}\frac{1}{A_i}\left((A_i+1){{\bf k}^\prime_{C_i}}+
{{\vvm p_j^\prime}_i} \right).
\end{equation}
Adding these two equations gives
\begin{equation}
\vvm K = \frac{A}{A+1}\frac{1}{A_i}\left((A_i+1) {\bf K}_{C_i}+{\vvm {\cal P}_j}_i\right).
\label{ap:11}
\end{equation}
This relation is the transformation prescription for $K$ from the cluster frame to 
the $A+1$ frame.
Solving for ${\bf K}_{C_i}$ in Eq.~(\ref{ap:11}) and plugging it into the
optical potential in the cluster frame of Eq.~(\ref{ap:7}) we are then able to 
express this potential completely using invariants or nucleon-nucleus $A+1$
variables,
\begin{eqnarray}
{U_{el}(\vvm q, \vvm K)}_{C_i} &=&\sum_{t=n,p}
\int d \vvm P \; d{\vvm {\cal P}_j}_i \; \rho_{corr} 
({\vvm {\cal P}_j}_i) \nonumber \\
&&\hat\tau_{0t}\left(\vvm q , \frac{1}{2}\left(\frac{A+1}{A}\vvm K - \vvm P\right),
{\cal{E}}\right)
\;\rho_i \left(\vvm P-\frac{A-1}{2A}\vvm q ,\vvm P+\frac{A-1}{2A}\vvm q \right).
\label{ap:7b}
\end{eqnarray}
This is Eq.~(\ref{eq:2.8.11c}) from Section II.C for the
cluster optical potential in the nucleon-nucleus
frame. 


\section{Correlation Density for the Cluster Approach}
\label{appendixb}

The cluster approach developed in this work uses a correlation density relating the
clusters, which is given in Eq.~(\ref{eq:c2}),
\begin{equation} 
\rho_{corr}({\vvm p_j}_1,{\vvm p^\prime_j}_1) 
 \equiv \int\prod_{l=2}^{N_c}d{{\vvm p^\prime_j}_l}
 \int\prod_{m=2}^{N_c}d{{\vvm p_j}_m} 
\langle \phi_A |\vvm {p^\prime_j}_1\vvm {p^\prime_j}_2 ...
\vvm {p^\prime_j}_{N_c}\rangle\;
\langle \vvm {p_j}_1\vvm {p_j}_2 ...
\vvm {p_j}_{N_c}
|\phi_A\rangle,
\end{equation}
This density correlates the momenta between the various clusters and elevates the
this approach beyond the independent single particle picture.
For the explicit derivation, let us start from  definition given in Eq.~(\ref{eq:c1})
\begin{equation}
{\vvm p_j}_i = \frac{1}{A}({A_s}_i \vvm p_i - A_i \vvm {p_s}_i).
\end{equation}
This definition is invariant from the frame of consideration. Thus it can be applied 
in the frame of the 
intrinsic density, where it describes the difference in momenta
between cluster $i$ and its analogous spectator particles, $\vvm {p_s}_i$, 
in the laboratory frame. In this same frame the total momentum between active
and spectator particles should add up to zero, at least before  the collision.
Thus, $\vvm p_i = - \vvm {p_s}_i$, and  therefore ${\vvm p_j}_i = \vvm p_i$ and
${{\vvm p_j}_i}^\prime = \vvm p_i - \vvm q$. Again, in the intrinsic frame of the 
nucleus one has
\begin{equation}
{\vvm {\cal P}_j}_i = \frac{{\vvm p_j}_i+ {\vvm p_j}^\prime_i}{2} =
{{\vvm p_j}_i} - \frac{\vvm q}{2}.
\end{equation}

For the $^6$He nucleus consisting of three clusters,  we 
can write the correlation density as
\begin{eqnarray}
\lefteqn{\rho_{corr}({\vvm p_j}_1,{\vvm p^\prime_j}_1) \equiv }\cr
&& \int d^3{\vvm p_{s_i}}_1 d (\vvm {\hat z} \cdot \vvm {\hat p_i})\;
d^3{{\vvm p^\prime_{s_i}}_1}\; d^3{\vvm p_{s_i}}_2\;
d^3{{\vvm p^\prime_{s_i}}_2}\; \; \Phi({\vvm p_{s_i}}_1)\; \Phi({\vvm p_{s_i}}_2)
\; \Phi({{\vvm p^\prime_{s_i}}_1})\; \Phi({{\vvm p^\prime_{s_i}}_2})\nonumber \\
&& f_{corr}(\vvm \Omega_{s_1}, \vvm \Omega_{s_2}) 
\delta(\vvm {p_i} - \vvm {p_{s_i}}_1 - \vvm i{p_{s_i}}_2)\;
\delta(\vvm {p_{s_i}}_1 - {\vvm {p^\prime_{s_i}}_1})\;
\delta(\vvm {p_{s_i}}_2 - {\vvm {p^\prime_{s_i}}_2}), \label{fcorr}
\end{eqnarray}
where the two spectator momenta are labeled ${\vvm p_{s_i}}_1$ and ${\vvm p_{s_i}}_2$.
The integration variables are over the momenta (before and after the scattering) of the
two spectators. We also integrate over the relationship of
the active particle's momentum to the quantization axis of $^6$He.
The first momentum conserving delta function preserves the c.m. momentum
(the last spectator therefore needs not be integrated over), the momenta of 
active and spectator particles must add up to zero. The
remaining two delta functions require that the  momentum of the spectator  
clusters do not
change. In practice, Eq.~(\ref{fcorr})  is for $^6$He a four dimensional integration.
In this work, the wave functions are the single particle wave functions, where the
momentum is converted to single particle form by simply dividing by the 
cluster mass. In one had a density defined using both in a cluster and a single
particle paradigm, then this formulation would allow for increased dynamical detail. The
angular correlation function, $f_{corr}$,
gives the correlation weighting assuming that
the spectator neutrons are in a given orbital shell, where
$(\vvm \Omega_{s_1}, \vvm \Omega_{s_2} )$ are the 
solid angles subtended by the spectator valence nucleons, in this case the $p\frac{3}{2}$
shell. 

For calculation the correlation density
we assume, that the two valence neutrons are in the
$p\frac{3}{2}$ shell, and  the total angular wave function can be written as
\begin{equation}
\psi_{p_\frac{3}{2}}= \frac{1}{2}(1-P_{12}){{\cal Y}_1^{\frac{3}{2}\frac{1}{2}}}
(\hat{\vvm\zeta_1}){{\cal Y}_1^{\frac{3}{2}\:-\frac{1}{2}}}
(\hat{\vvm\zeta_2})
+ \frac{1}{2}(1-P_{12}){{\cal Y}_1^{\frac{3}{2}\frac{3}{2}}}
(\hat{\vvm\zeta_1}){{\cal Y}_1^{\frac{3}{2}\:-\frac{3}{2}}}
(\hat{\vvm\zeta_2}),\label{fcorr2}
\end{equation}
where ${\cal Y}_l^{j\;m_j}(\hat{\vvm\zeta})$
are the traditional spin spherical harmonics and $\vvm \zeta$ is the single 
particle intrinsic momentum. The antisymmetric of the wave function with respect to
the two neutrons is given by the operator
$(1-P_{12})$.
The correlation function, $f_{corr}(\vvm \Omega_{s_1}, \vvm \Omega_{s_2})$, 
defined in Eq.~(\ref{fcorr}) can be with the help of Eq.~(\ref{fcorr2}) defined as
\begin{equation}
f_{corr} = \psi_{p_\frac{3}{2}}^*(\hat{\vvm\zeta_1},\hat{\vvm\zeta_2})
\psi_{p_\frac{3}{2}}(\hat{\vvm\zeta_1},\hat{\vvm\zeta_2}).
\end{equation}
This defines the angular probability for the two  spectator neutrons when
the alpha core is active. This local correlation function is also used
as the approximate 
probability when a neutron is the active particle, i.e. when the full calculation
is in fact off-shell. The alpha core has an unweighted angular distribution,
approximated by the COSMA density, as being completely in the $s$-orbital.

Explicitly inserting the spherical harmonics into Eq.~(\ref{fcorr2}), we obtain
\begin{eqnarray}
f_{corr} &=& \frac{1}{32 \pi^2}{\big ( 2 \cos\theta_1\cos\theta_2 - 
\sin\theta_1\sin\theta_2 \cos(\phi_1-\phi_2)\big ) }^2 \nonumber \\
&+& \frac{1}{32 \pi^2}\big (\sin^2\theta_1\cos^2\theta_2 + 
\cos^2\theta_1\sin^2\theta_2 - \sin 2\theta_1
\sin 2\theta_2\cos(\phi_1-\phi_2)\big ).
\end{eqnarray}
The first term is the result if the total spin projection of
the two neutrons added up to zero and
the second term is due to  the total spin projection being one.

Once the four dimensional integral for the  cluster correlation,
Eq.~(\ref{fcorr}), is calculated, it is normalized to one and then 
used to augment the definition of the optical potential. De facto, it
 constrains the momentum
of the  c.m. of the active cluster  relative to the spectators.


\begin{acknowledgments}
This work was performed in part under the
auspices of the U.~S.  Department of Energy under contract
No. DE-FG02-93ER40756 with Ohio University and
under contract No. DE-SC0004084 (TORUS Collaboration).  S.P.W. thanks the
Institute of Nuclear and Particle Physics (INPP) and the Department of
Physics and Astronomy at Ohio University for their hospitality and
support during his sabbatical stay.
\end{acknowledgments}



%

\clearpage

\noindent

\end{document}